\documentclass[prl,aps,amsmath,superscriptaddress,twocolumn,color,dvipdfmx]{revtex4-2}

\usepackage{graphicx}% Include figure files
\usepackage{dcolumn}% Align table columns on decimal point
\usepackage{bm}% bold math
\usepackage[T1]{fontenc}

\begin{document}

\title{Quantum droplets with magnetic vortices in spinor dipolar
  Bose-Einstein condensates}

\author{Shaoxiong Li}
\affiliation{Department of Engineering Science, University of Electro-Communications, Tokyo 182-8585, Japan
}
\author{Hiroki Saito}
\affiliation{Department of Engineering Science, University of Electro-Communications, Tokyo 182-8585, Japan
}

\date{\today}

\begin{abstract}
Motivated by the recent experimental realization of a Bose-Einstein
condensate (BEC) of europium atoms, we investigate the self-bound
droplet state of a europium BEC with spin degrees of freedom.
Under a sufficiently weak magnetic field, the droplet has a torus
shape with circulating spin vectors, which is referred to as a
magnetic vortex.
The ground state transforms from the torus to cigar shape through
bistability with an increase in the magnetic field.
Dynamical change of the magnetic field causes the torus to rotate due
to the Einstein-de Haas effect.
The magnetic vortices form a supersolid in a confined system.
\end{abstract}

\maketitle

A magnetic flux-closure structure is a stable configuration of a
ferromagnetic material, in which magnetization vectors form closed
loops to reduce the magnetostatic energy.
This structure can be observed in a ferromagnetic material with
multiple magnetic domains below the Curie temperature~\cite{Kittel}.
The magnetic flux-closure structure has also been realized within
nanoscale particles~\cite{Tonomura, Cowburn, Shinjo, Rothman, SPLi,
  Wacho, Hytch, Jia, Lewis}, in which the magnetization vectors
circulate along a toroidal loop.
This state is referred to as a magnetic vortex.
The nanoparticles with magnetic vortices can be used for, e.g., data
storage~\cite{Cowburn2}, cancer therapy~\cite{Kim}, and neuromorphic
computing~\cite{Torrejon}.
Such an isolated object with a magnetic vortex has so far been
restricted to solid materials.
Is it possible to produce a liquid or gas analogue of this state of
matter, i.e., a self-bound droplet with a magnetic vortex?
Although permanent-magnetic liquid droplets have been produced
recently~\cite{Liu}, the magnetic-vortex structure has not been
observed.

Here we propose a self-bound superfluid droplet that contains a
magnetic vortex.
Self-bound states of Bose-Einstein condensates (BECs) have attracted
much interest recently and are referred to as quantum
droplets~\cite{Petrov}.
In a quantum droplet, the attractive mean-field interaction balances
with the repulsive beyond-mean-field effect~\cite{LHY}, which
stabilizes the system against collapse and expansion in free space.
This novel state of matter was realized in a BEC with magnetic
dipole-dipole interaction
(DDI)~\cite{Kadau,4schmitt2016self,5PhysRevLett.116.215301,6PhysRevX.6.041039,7PhysRevLett.120.160402}
and a Bose-Bose mixture~\cite{Cabrera, Semeghini, Cheiney},
and various theoretical studies have been performed on these
systems~\cite{waechtler2016,waechtler2016b,Saito,Bisset,macia2016droplets,8PhysRevA.96.053630,9PhysRevLett.119.255302,cidrim2018vortices,13PhysRevA.98.063602,14PhysRevLett.122.193902,15PhysRevResearch.1.033088,16PhysRevLett.122.090401,17PhysRevLett.123.133901,oldziejewski2020strongly,Lee,Yogurt,30bisset2021,smith2021quantum,20PhysRevLett.130.043401}.
However, in the experiments of quantum droplets with DDI to date,
the magnetization of the system has been frozen into the direction of
the strong external magnetic field.
If the external magnetic field is suppressed sufficiently, spin
degrees of freedom in a dipolar BEC are liberated, which allows for a
spinor dipolar
BEC~\cite{22PhysRevLett.93.040403,Kawaguchi_EdH,Santos,Diener,Yi,Kawaguchi2,26PhysRevLett.98.110406,Takahashi,Gawryluk,Sun,Yasunaga,hoshi2010,Swislocki,Huhtamaki,Deure,kudo2010hydrodynamic,Simula,Gaw11,Pasquiou,Eto,Marti,Kunimi,Oshima,Borgh,liao2017anisotropic,lepoutre2018spin}.
However, self-bound quantum droplets of spinor dipolar BECs have
not yet been studied.

In this paper, we will show that there exists a stable self-bound
droplet of a spinor dipolar BEC that contains a magnetic vortex with a
torus-shaped density distribution.
This state is stable under a weak external magnetic field.
As the the magnetic field is increased, the ground state changes from
the magnetic vortex state to the well-known cigar-shaped droplet, and
these two states exhibit bistability.
If the external magnetic field is suddenly changed, the torus-shaped
droplet begins to rotate to conserve the total angular momentum, which
resembles the Einstein-de Haas effect.
The ground state of a confined system exhibits periodic alignment of
the torus-shaped droplets, which can be regarded as a supersolid.

In the present study, we restrict ourselves to the BEC of
$^{151}\text{Eu}$ atoms, which was recently realized
experimentally~\cite{33PhysRevLett.129.223401}.
A peculiar feature of $^{151}\text{Eu}$ is its wide range of the
hyperfine spin ($F = 1, \cdots, 6$) with small spin-dependent contact
interactions, which could be smaller than the DDI.
The spin state in this system is therefore determined mainly by the
DDI under a weak external magnetic field, and the DDI-dominant spinor
dipolar phenomena can be investigated.
To observe such phenomena, the magnetic Feshbach resonance cannot be
used to tune the contact interaction.
Although the $s$-wave scattering length of the $F = 6$ hyperfine state
measured in Ref.~\cite{33PhysRevLett.129.223401} does not satisfy the
condition for the droplet formation, the scattering lengths for the
other hyperfine spins $F$ and those for $^{153}\text{Eu}$ are unknown,
for which the spinor dipolar droplet may be possible.
Furthermore, the contact interaction may be controllable using
microwave-induced Feshbach resonance~\cite{34PhysRevA.81.041603},
which enables the formation of the spinor dipolar droplet.

We consider a BEC of $^{151}\text{Eu}$ atoms with hyperfine spin $F$
at zero temperature using the beyond-mean-field
approximation~\cite{waechtler2016, waechtler2016b}.
The total energy consists of five terms,
$
E=E_{\rm kin}+E_{s}+E_{\rm ddi}+E_{\rm LHY}+E_{B}.
$
The kinetic energy is given by
$E_{\rm kin} = \hbar^2 / (2 M) \sum_m \int d\bm{r}
\left|\nabla \psi_m(\bm{r})\right|^2$,
where $\psi_m(\bm{r})$ is the macroscopic wave function for the
magnetic sublevels $m=-F,-F+1,\cdots, F$, and $M$ is the mass of an
atom.
The wave function is normalized as $\sum_m \int\left|
\psi_m(\bm{r}) \right|^2d\bm{r}=N$, where $N$ is the
total number of atoms.
The spin-independent contact interaction has the form
$E_{s} = 2\pi\hbar^2 a_s M^{-1} \int\rho^2(\bm{r})d\bm{r}$,
where $a_s$ is the spin-independent $s$-wave scattering length and
$\rho(\bm{r})=\sum_m\left|\psi_m(\bm{r})\right|^2$ is the
total density.
The DDI energy is given by
$E_{\rm ddi} = \mu_0 (g\mu_B)^2 / (8\pi) \int d\bm{r}d\bm{r}' \{
\bm{f}(\bm{r}) \cdot \bm{f}(\bm{r}') - 3 [\bm{f}(\bm{r}) \cdot \bm{e}]
   [\bm{f} (\bm{r}') \cdot \bm{e}] \} / |\bm{r}-\bm{r}'|^3$,
where $\mu_0$ is the magnetic permeability of the vacuum, $g$ is the
hyperfine $g$ factor, $\mu_B$ is the Bohr magneton,
$\bm{f}(\bm{r}) = \sum_{mm'}\psi_m^*(\bm{r})(\bm{S})_{mm'}\psi_{m'}(\bm{r})$
with $\bm{S}$ being the spin matrix, and
$\bm{e}=(\bm{r}-\bm{r}') / |\bm{r} - \bm{r}'|$.
The relative strength of the DDI is characterized by
$\varepsilon_{\rm dd} = a_{\rm dd} / a_s$, where $a_{\rm dd} = \mu_0
\mu^2 M / (12\pi\hbar^2)$ is the dipolar length.
The magnetic moment $\mu = g \mu_B F$ and the dipolar length
$a_{\rm dd}$ for each spin $F$ of $^{151}\text{Eu}$ are given in the
Supplemental Material.
The spin distribution is mainly determined by the DDI, if $a_{\rm dd}$
is much larger than the spin-dependent scattering lengths that
consist of the differences $\Delta a$ among the scattering lengths
$a_{0, 2, \cdots, 2F}$ in collisional spin channels.
Since the values of $\Delta a$ are predicted to be relatively small
for the europium atoms~\cite{33PhysRevLett.129.223401, 35JCP,
36PhysRevA.81.022701}, we ignore the spin-dependent contact
interaction.

As will be confirmed numerically, the spin state is almost fully
polarized in the droplet, and we can use the Lee-Huang-Yang (LHY)
correction for the fully polarized dipolar BEC.
Under the local density approximation, the LHY correction is written
as~\cite{lima2011, lima2012, waechtler2016, waechtler2016b} 
\begin{equation} \label{LHY}
E_{\rm LHY} = \frac{2}{5} \frac{32}{3\sqrt{\pi}} \frac{4\pi \hbar^2}{M} a^{5/2} \chi(\varepsilon_{\rm dd}) \int \rho^{5/2}(\bm{r}) d\bm{r},
\end{equation}
where $\chi(\varepsilon_{\rm dd})$ is the real part of
$\int_0^\pi d\theta \sin\theta [1 + \varepsilon_{\rm dd}
  (3\cos^2\theta - 1)]^{5/2} / 2$.
In the presence of an external magnetic field $\bm{B}(\bm{r})$, the
linear Zeeman energy has the form
$E_{B} = -g \mu_B \int \bm{f}(\bm{r}) \cdot \bm{B}(\bm{r})d\bm{r}$.
The ratio of the quadratic Zeeman energy to the linear Zeeman energy
is estimated to be $\mu B / \Delta_{\rm hf} \sim 10^{-4}$ at most for
the present magnetic field $\sim 0.1$ mG, and the quadratic Zeeman
energy can be neglected even for the relatively small hyperfine
splitting $\Delta_{\rm hf} / \hbar \sim 100$ MHz of a europium
atom~\cite{43jin2002}.

The Gross-Pitaevskii (GP) equation is given by the functional
derivative of the total energy as
$i \hbar \partial \psi_m / \partial t = \delta E / \delta \psi_m^*$.
To obtain the ground state or metastable state, the GP equation is
propagated in imaginary time, in which $i$ on the left-hand side of
the GP equation is replaced with $-1$.
The GP equation is numerically solved using the pseudospectral method
with typical spatial and time steps $dx \sim 0.01$ $\mu{\rm m}$ and
$dt \sim 0.1$ $\mu{\rm s}$.
\iffalse
\begin{table}
    \centering
    \begin{tabular}{|c|c|c|c|c|c|c|}
    \hline
        $F$ & 1 & 2 & 3 & 4 & 5 & 6\\
    \hline
        $\mu/\mu_B$& 9/2 & 13/3 & 19/4 & 27/5 & 37/6 & 7\\
    \hline
        $a_{dd}/a_B$ & 24.72 & 22.92 & 27.54 & 35.60 & 46.42 & 59.82\\
    \hline
    \end{tabular}
    \caption{Magnetic moment $\mu$ in unit of the Bohr magneton $\mu_B$ and the dipolar lengths $a_{dd}$ in unit of the Bohr radius $a_B$, for $^{151}\text{Eu}$ with hyperfine spin $F$. }
    \label{tab:my_label}
\end{table}
\fi
\begin{figure}
\centering
\includegraphics[width=7.5cm]{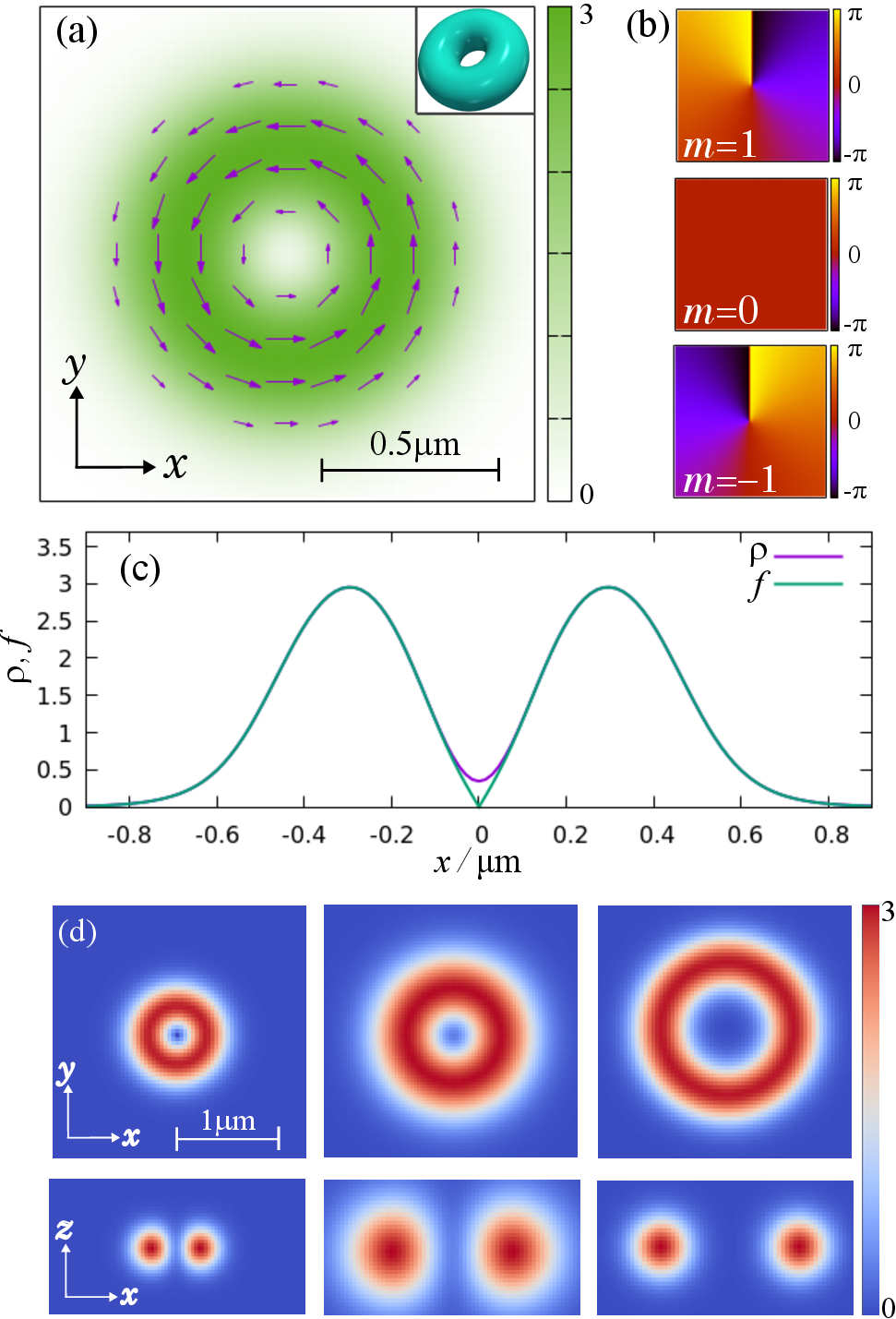}
\caption{
  Self-bound ground state with a magnetic vortex for $B = 0$.
  The system has translational and rotational symmetry, and the
  origin and $z$ axis are taken as the center and symmetry axis of the
  torus, respectively.
  (a) Density distribution on the $z = 0$ plane for $F=1$, $N =15000$,
  and $\varepsilon_{\rm dd} = 1.2$.
  The arrows represent the magnetization $\bm{f}$.
  %The size of the panel is $1.4 \times 1.4$ $\mu{\rm m}$.
  The inset shows the isodensity surface.
  (b) Phase distributions of components $m = 1, 0, -1$ on the $z = 0$
  plane.
  (c) Distributions of $\rho$ and $|\bm{f}|$ along the $x$ axis. 
  (d) Density distributions on the $z = 0$ plane (upper panels) and $y
  = 0$ plane (lower panels) for $(F, N, \varepsilon_{\rm dd}) = 
  (1, 15000, 1.2)$, $(1, 80000, 1.2)$, and $(6, 15000, 1.3)$.
  %The size is $2.5 \times 2.5$ $\mu{\rm m}$ and $2.5 \times 1.25$
  %$\mu{\rm m}$ for the upper and lower panels.
  The unit of density in (a), (c), and (d) is $N \mu{\rm m}^{-3}$.
}
\end{figure}

First, we consider the case in which the external magnetic field $B$ is zero.
Figure~1(a) shows a typical ground state, which has a torus shape, in
contrast to the usual cigar-shaped droplet in a strong magnetic
field~\cite{Kadau,4schmitt2016self,5PhysRevLett.116.215301,6PhysRevX.6.041039,7PhysRevLett.120.160402}. 
The magnetization vectors $\bm{f}$ circulate along the torus, as shown
by the arrows in Fig.~1(a).
Although such a magnetic vortex state has already been proposed for a
trapped spinor dipolar BEC~\cite{Kawaguchi2,Takahashi}, this is the first example of a
self-bound droplet of a fluid containing a magnetic vortex.

In the LHY energy in Eq.~(\ref{LHY}), we assumed that the spin state
is fully polarized, and here we examine the validity of this
assumption.
Figure~1(c) shows the distributions of the atomic density $\rho$ and
magnetization density $|\bm{f}|$, which indicates that the spin is
almost fully polarized, i.e., $|\bm{f}| / \rho \simeq F = 1$, except
near the center.
This result justifies the use of Eq.~(\ref{LHY}), since the LHY
correction is important only in the high-density region to counteract
the collapse.
The central hole of the torus is mainly occupied by the $m = 0$
component, since the $m \neq 0$ components have the topological
defects at the center, as shown in Fig.~1(b).

Figure~1(d) shows the parameter dependence of the density profile.
The size of the droplet increases with the number of atoms $N$, while
the aspect ratio between the major and minor radii appears almost
unchanged.
For larger spin $F=6$, on the other hand, the hole of the torus is
enlarged and the aspect ratio is significantly changed.
This is due to the kinetic energy that arises from the spin winding,
which is proportional to $F$.
Such behavior is analyzed in the Supplemental Material using the
variational method.

We employ the variational wave function as,
\begin{equation} \label{psiv}
  \bm{\Psi}_v(\bm{r}) = \sqrt{\rho_v(r, z)} e^{-i {S}_z \phi}
  \bm{\zeta}^{(y)},
\end{equation}
where $(r, \phi, z)$ is the cylindrical coordinate, and
$\bm{\zeta}^{(y)}$ represents the spin state fully polarized in the
$y$ direction with $\sum_m |\zeta_m^{(y)}|^2 = 1$.
The $z$ axis is taken as the symmetry axis of the torus.
The matrix $e^{-i {S}_z \phi}$ rotates the spin vector to make a
magnetic vortex.
We propose a torus-shaped variational density as
\begin{equation} \label{rhov}
\rho_v(r, z) = \frac{N}{\pi^{3/2} \sigma_r^{2\lambda + 2} \sigma_z
  \Gamma(\lambda + 1)} r^\lambda
e^{-\frac{r^2}{\sigma_r^2} - \frac{z^2}{\sigma_z^2}},
\end{equation}
where $\sigma_r > 0$, $\sigma_z > 0$, and $\lambda > 0$ are
variational parameters and $\Gamma$ is the gamma function.
The variational energy for Eqs.~(\ref{psiv}) and (\ref{rhov}) is
derived in the Supplemental Material, which is numerically minimized
with respect to the variational parameters using the Newton-Raphson
method.
%We note that $E_{\rm ddi} / E_s = -\varepsilon_{\rm dd}$ holds for
%Eqs.~(\ref{psiv}) and (\ref{rhov}), and therefore
%$\varepsilon_{\rm dd} > 1$ is the necessary condition for the droplet
%to be bound by the attractive part of the DDI.

\begin{figure}
\includegraphics[width=7.5cm]{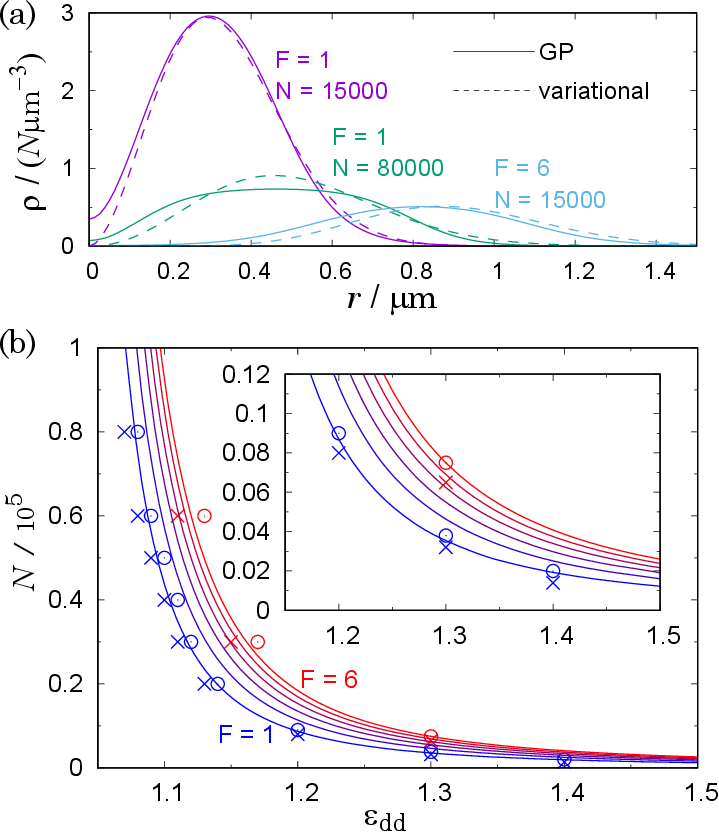}
\caption{
(a) Density distribution $\rho(r, z = 0)$ obtained by the
imaginary-time evolution of the GP equation (solid lines) and the
variational method (dashed lines) for $(F, N, \varepsilon_{\rm dd}) =
(1, 15000, 1.2)$, $(1, 80000, 1.2)$, and $(6, 15000, 1.3)$.
(b) Lines: the critical number of atoms above which the
droplet is stable, obtained by the variational method.
The six lines represent $F = 1$, $\cdots$, 6 from left to right.
The circles (crosses) indicate that the droplet is stable (unstable)
for $F = 1$ (blue or dark gray) and $F = 6$ (red or light gray),
obtained by the GP equation.
The inset shows a magnification of the main panel.
}
\end{figure}

Figure 2(a) compares the density distributions $\rho(r, z = 0)$
obtained by the GP equation and by the variational method.
The two distributions agree well with each other.
For $N = 80000$, the distribution of the GP result becomes broader
than that of the variational method, since the flat-top tendency of a
large droplet is not taken into account in Eq.~(\ref{rhov}).
The density at $r = 0$ must vanish for the fully-polarized assumption
in Eq.~(\ref{psiv}), whereas the center is slightly occupied for the
GP results.
Figure~2(b) shows the critical number of atoms above which the droplet
is stable, where the lines are obtained by the variational method and
the plots by the GP equation.
The variational method can predict the critical number of atoms very
well, which facilitates the study of this system, because the
numerical cost for the GP simulation is much higher than that for the
variational method, especially near the critical line of the stability
where convergence of the imaginary-time evolution is slow.

\begin{figure}
\centering
\includegraphics[width=7.5cm]{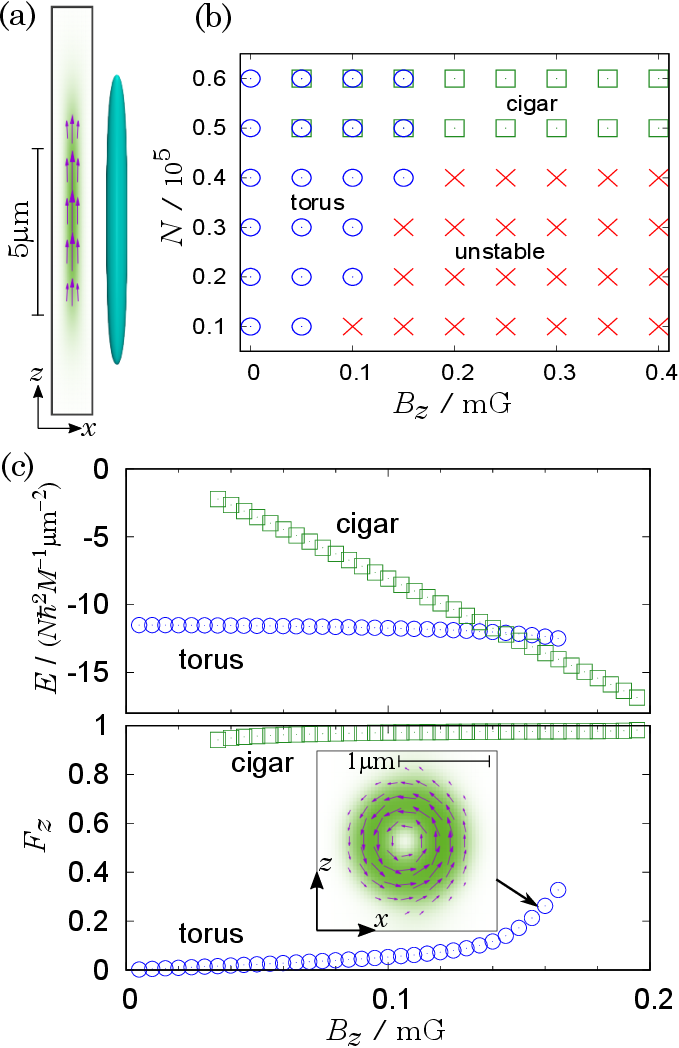}
\caption{
  Effects of magnetic field $B_z$ obtained by the GP equation
  for $F = 1$ and $\varepsilon_{\rm dd} = 1.2$.
  (a) Cross-sectional $\rho$ and $\bm{f}$ distributions and isodensity
  surface of the ground state for $N = 50000$ and $B_z = 0.2$ mG.
  %The size of the panel is $1.2 \times 12$ $\mu{\rm m}$.
  (b) Stability diagram.
  Parameter sets marked by circles and squares respectively indicate
  that the torus-shaped and cigar-shaped droplets are stable or
  metastable.
  (c) $B_z$ dependence of the energy $E$ and averaged magnetization
  $F_z$ of the torus-shaped and cigar-shaped droplets with $N = 50000$.
  The inset shows $\rho$ and $\bm{f}$ distributions of the
  torus-shaped droplet for $B_z = 0.16$ mG, where the cross section is
  taken for the symmetry plane ($y = 0$ plane).
  %The size of the inset is $2 \times 2$ $\mu{\rm m}$.
}
\end{figure}

We next examine the effect of the external magnetic field applied in
the $z$ direction.
Figure 3(a) shows a typical ground state for a large magnetic field
$B_z$, where the droplet has a cigar-shape, as observed
experimentally.
The spin is almost polarized in the $z$ direction, whereas it is
slightly tilted around both edges of the cigar-shape and exhibits the
flower-like structure~\cite{Kawaguchi2}.
Figure 3(b) shows the stability diagram with respect to $N$ and $B_z$.
There is a critical magnetic field, above which the torus-shaped
droplet becomes unstable, whereas the cigar-shaped droplet becomes
unstable below some critical magnetic field.
In Fig.~3(b), there is a bistability region in which both torus-shaped
and cigar-shaped droplets are stationary (both circles and squares are
marked).
Figure~3(c) reveals the bistability with plots of the energy $E$ and
the averaged magnetization in the $z$ direction $F_z = \int f_z
d\bm{r} / N$ for both droplets.
The bistability ranges from $B_z\simeq 0.03$ to $\simeq 0.17$ mG, and
the energies of the two droplets cross at $B_z \simeq 0.14$ mG.
The direction of the torus-shaped droplet is fixed by the magnetic
field in such a way that the toroidal plane is parallel to the $z$
direction, as shown in the inset of Fig.~3(c).
In this inset, the right-hand side of the torus becomes slightly
thicker than the left-hand side, which results in the increase of
$F_z$.

\begin{figure}
\centering
\includegraphics[width=7.5cm]{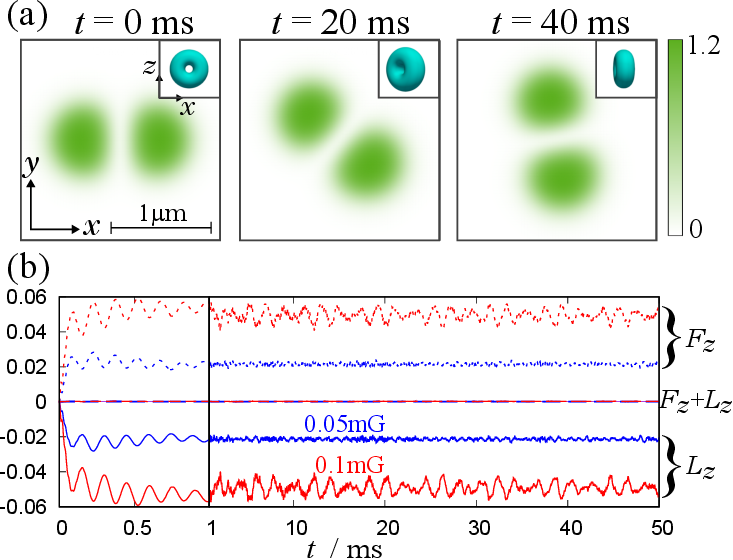}
\caption{
  Einstein-de Haas effect of the torus-shaped droplet for $F = 1$,
  $\varepsilon_{\rm dd} = 1.2$, and $N =15000$.
  The initial state is the ground state for zero magnetic field (the
  same state as Fig.~1(a), except that the symmetry axis of the torus
  is taken to be the $y$ direction).
  (a) Time evolution of the density distribution on the $z = 0$ plane
  (main panels) and isodensity surface observed from the $-y$
  direction (insets), where the magnetic field $B_z = 0.1$ mG is
  switched on at $t = 0$.
  %The size of each panel is $2 \times 2$ $\mu{\rm m}$.
  The unit of density is $N \mu{\rm m}^{-3}$.
  See the Supplemental Material for the movie of the dynamics.
  (b) Time evolution of the orbital angular momentum $L_z$ (solid
  lines), spin angular momentum $F_z$ (dotted lines), and total
  angular momentum $F_z + L_z$ (dashed lines) for $B_z = 0.05$
  (blue or dark gray) and 0.1 mG (red or light gray).
  The first 1 ms is magnified.
}
\end{figure}

The increase of the magnetization $F_z$ with the magnetic field $B_z$
implies the emergence of the Einstein-de Haas effect~\cite{Richardson,
 EdH}; if the applied magnetic field $B_z$ is increased dynamically,
the spin angular momentum $F_z$ will increase, which must be
accompanied by a decrease in the orbital angular momentum $L_z = -i
\int \sum_m  d\bm{r} \psi_m^* (x \partial_y - y \partial_x) \psi_m^*$
to conserve the total angular momentum.
Figure~4 demonstrates the dynamics of the Einstein-de Haas effect,
where the initial droplet state is prepared for zero magnetic field
with the symmetry axis in the $y$ direction, and the magnetic field
$B_z$ is turned on at $t = 0$.
As expected, the droplet begins to rotate around the $z$ axis, where
the total angular momentum $F_z + L_z$ is maintained to be zero.
We note that such mechanical rotation of the torus is a clearer
manifestation of the Einstein-de Haas effect than that in the trapped
system~\cite{Kawaguchi_EdH}.

\begin{figure}
\centering
\includegraphics[width=7.5cm]{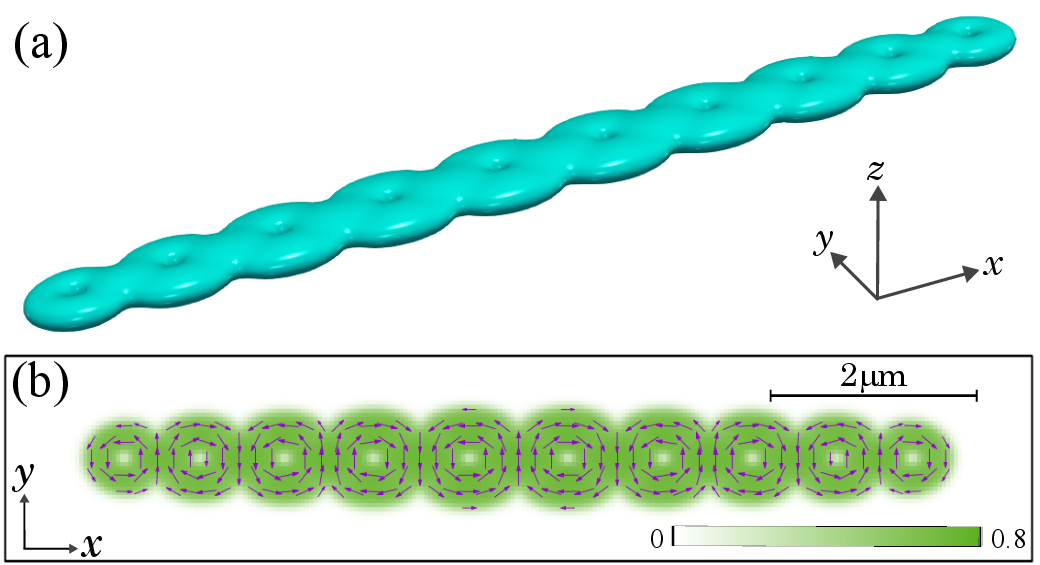}
\caption{
  Ground state for $F = 1$, $N =4 \times 10^5$, $\varepsilon_{\rm dd}
  = 1.4$, and $B_z = 0$ in a harmonic potential $M (\omega_x^2 x^2 +
  \omega_y^2 y^2 + \omega_z^2 z^2) / 2$ with $(\omega_x, \omega_y,
  \omega_z) = 2\pi \times (100, 1500, 6000)$ Hz.
  (a) Isodensity surface and (b) distributions of the density $\rho$
  and magnetization $\bm{f}$ on the $z = 0$ plane.
  The unit of density is $N \mu{\rm m}^{-3}$.
  %The size of the panel in (b) is $10 \times 2$ $\mu{\rm m}$.
}
\end{figure}

Next we consider the ground state of the system confined in a
``surfboard-shaped'' trap~\cite{StamperKurn}, where the trap is
weak, moderate, and tight in the $x$, $y$, and $z$ directions,
respectively.
Figure~5 shows the density and spin distributions of the ground state,
in which multiple droplets with magnetic vortices are aligned along
the $x$ axis with alternate circulations of the magnetic vortices.
This state can be called a one-dimensional supersolid, because the
ground state has the (quasi)periodicity that breaks the
(quasi)translation symmetry (in the $x$ direction), while each droplet
is connected with adjacent droplets, which enables superflow between
them.
We note that the dipolar supersolid of cigar-shaped
droplets~\cite{Tanzi19, Bottcher19, Chomaz19, Tanzi19_2, Guo19,
  Natale19, Tanzi21, Ilzhofer21, Hertkorn21, Sohmen21, Norcia21,
  Bland22, Biagioni22, Norcia22, Sanchez23}
also requires confinement in one or two directions, i.e., restricted
geometry is required for the dipolar BEC to split into multiple
droplets.
If the trap in the $y$ or $z$ direction in Fig.~5 is removed, the
ground state becomes a large single droplet with a magnetic vortex. 

The torus-shaped droplet may be generated experimentally by the
following procedure.
First, the condensate atoms are prepared in the $|F, m = 0\rangle$
hyperfine state in an optical trap.
Since this nonmagnetized state is dynamically and energetically
unstable, spontaneous magnetization occurs~\cite{hoshi2010}.
If the BEC is confined to the expected droplet size, the spontaneous
magnetization will form the magnetic vortex with the lowest
energy~\cite{Kawaguchi2}.
After some relaxation time, the optical trap is switched off, which
results in the self-bound torus-shaped droplet.
When the hyperfine spins with higher energies ($F \neq 6$) are used,
the above experimental procedure must be accomplished within the
lifetime due to hyperfine exchanging collisions, which has not been
measured for a europium BEC.

To summarize, we have investigated a self-bound droplet in a spinor
dipolar BEC.
For a large number of atoms, large $\varepsilon_{\rm dd}$, and a small
magnetic field, there exists a stable self-bound torus-shaped droplet
that contains a magnetic vortex (Fig.~1).
For some range of the magnetic field, the system exhibits bistability
between the torus and cigar-shaped droplets (Fig.~3).
The torus-shaped droplet exhibits the Einstein-de Haas effect by the
change of the applied magnetic field (Fig.~4).
Multiple droplets form a supersolid in a surfboard-shaped trap
(Fig.~5).

This work was supported by JSPS KAKENHI Grant No. JP23K03276.


\begin{thebibliography}{99}

\bibitem{Kittel}
  For example, C. Kittel, {\it Introduction to Solid State Physics},
  8th ed., Chap. 12 (John Wiley \& Sons, New Jersey, 2005).

\bibitem{Tonomura}
  A. Tonomura, T. Matsuda, J. Endo, T. Arii, and K. Mihama,
  Direct observation of fine structure of magnetic domain walls by
  electron holography,
  Phys. Rev. Lett. \textbf{44}, 1430 (1980).

\bibitem{Cowburn}
  R. P. Cowburn, D. K. Koltsolv, A. O. Adeyeye, M. E. Welland, and
  D. M. Tricker,
  Single-domain circular nanomagnets,
  Phys. Rev. Lett. \textbf{83}, 1042 (1999).

\bibitem{Shinjo}
  T. Shinjo, T. Okuno, R. Hassdorf, K. Shigeto, and T. Ono,
  Magnetic vortex core observation in circular dots of permalloy,
  Science \textbf{289}, 930 (2000).

\bibitem{Rothman}
  J. Rothman, M. Kl\"aui, L. Lopez-Diaz, C. A. F. Vaz, A. Bleloch,
  J. A. C. Bland, Z. Cui, and R. Speaks,
  Observation of a bi-domain state and nucleation free switching in
  mesoscopic ring magnets,
  Phys. Rev. Lett. \textbf{86}, 1098 (2001).

\bibitem{SPLi}
  S. P. Li, D. Peyrade, M. Natali, A. Lebib, Y. Chen, U. Ebels,
  L. D. Buda, and K. Ounadjela,
  Flux closure structures in cobalt rings,
  Phys. Rev. Lett. \textbf{86}, 1102 (2001).

\bibitem{Wacho}
  A. Wachowiak, J. Wiebe, M. Bode, O. Pietzsch, M. Morgenstern, and
  R. Wiesendanger,
  Direct observation of internal spin structure of magnetic vortex
  cores,
  Science \textbf{298}, 577 (2002).

\bibitem{Hytch}
  M. J. H\"ytch, R. E. Dunin-Brokowski, M. R. Scheinfein, J. Moulin,
  C. Duhamel, F. Mazaleyrat, and Y. Champion,
  Vortex flux channeling in magnetic nanoparticle chains,
  Phys. Rev. Lett. \textbf{91}, 257207 (2003).

\bibitem{Jia}
  C.-J. Jia, L.-D. Sun, F. Luo, X.-D. Han, L. J. Heyderman, Z.-G. Yan,
  C.-H. Yan, K. Zheng, Z. Zhang, M. Takano, N. Hayashi, M. Eltschka,
  M. Kl\"aui, U. R\"udiger, T. Kasama, L. Cervera-Gontard,
  R. E. Dunin-Borkowski, G. Tzvetkov, and J. Raabe,
  Large-scale synthesis of single-crystalline iron oxide magnetic
  nanorings,
  J. Am. Chem. Soc. \textbf{130}, 16968 (2008).

\bibitem{Lewis}
  G. R. Lewis, J. C. Loudon, R. Tovey, Y.-H. Chen, A. P. Roberts,
  R. J. Harrison, P. A. Midgley, and E. Ringe,
  Magnetic vortex states in toroidal iron oxide nanoparticles:
  combining micromagnetics with tomography,
  Nano Lett. \textbf{20}, 7405 (2020).

\bibitem{Cowburn2}
  R. P. Cowburn,
  Magnetic nanodots for device applications,
  J. Magn. Magn. Mater. \textbf{242-245}, 505 (2002).

\bibitem{Kim}
  D.-H. Kim, E. A. Rozhkova, I. V. Ulasov, S. D. Bader, T. Rajh,
  M. S. Lesniak, and V. Novosad,
  Biofunctionalized magnetic-vortex microdiscs for targeted
  cancer-cell destruction,
  Nat. Mater. \textbf{9}, 165 (2009).

\bibitem{Torrejon}
  J. Torrejon, M. Riou, F. A. Araujo, S. Tsunegi, G. Khalsa,
  D. Querlioz, P. Bortolotti, V. Cros, K. Yakushiji, A. Fukushima,
  H. Kubota, S. Yuasa, M. D. Stiles, and J. Grollier,
  Neuromorphic computing with nanoscale spintronic oscillators,
  Nature (London) \textbf{547}, 428 (2017).

\bibitem{Liu}
  X. Liu, N. Kent, A. Ceballos, R. Streubel, Y. Jiang, Y. Chai,
  P. Y. Kim, J. Forth, F. Hellman, S. Shi, D. Wang, B. A. Helms,
  P. D. Ashby, P. Fischer, and T. P. Russell,
  Reconfigurable ferromagnetic liquid droplets,
  Science \textbf{365}, 264 (2019).

\bibitem{Petrov}
  D. S. Petrov,
  Quantum mechanical stabilization of a collapsing Bose-Bose mixture,
  Phys. Rev. Lett. \textbf{115}, 155302 (2015).

\bibitem{LHY}
  T. D. Lee, K. Huang, and C. N. Yang,
  Eigenvalues and eigenfunctions of a Bose system of hard spheres and
  its low-temperature properties,
  Phys. Rev. \textbf{106}, 1135 (1957).

\bibitem{Kadau}
  H. Kadau, M. Schmitt, M. Wenzel, C. Wink, T. Maier, I. Ferrier-Barbut,
  and T. Pfau,
  Observing the Rosensweig instability of a quantum ferrofluid,
  Nature (London) \textbf{530}, 194 (2016).

\bibitem{4schmitt2016self}
  M. Schmitt, M. Wenzel, F. B\"ottcher, I. Ferrier-Barbut, and T. Pfau,
  Self-bound droplets of a dilute magnetic quantum liquid,
  Nature (London) \textbf{539}, 259 (2016).

\bibitem{5PhysRevLett.116.215301}
  I. Ferrier-Barbut, H. Kadau, M. Schmitt, M. Wenzel, and T. Pfau,
  Observation of quantum droplets in a strongly dipolar Bose gas,
  Phys. Rev. Lett. \textbf{116}, 215301 (2016).

\bibitem{6PhysRevX.6.041039}
  L. Chomaz, S. Baier, D. Petter, M. J. Mark, F. W\"achtler, L. Santos,
  and F. Ferlaino,
  Quantum-fluctuation-driven crossover from a dilute Bose-Einstein
  condensate to a macrodroplet in a dipolar quantum fluid,
  Phys. Rev. X \textbf{6}, 041039 (2016).

\bibitem{7PhysRevLett.120.160402}
  I. Ferrier-Barbut, M. Wenzel, F. B\"ottcher, T. Langen, M. Isoard,
  S. Stringari, and T. Pfau,
  Scissors mode of dipolar quantum droplets of dysprosium atoms,
  Phys. Rev. Lett. \textbf{120}, 160402 (2018).

\bibitem{Cabrera}
  C. R. Cabrera, L. Tanzi, J. Sanz, B. Naylor, P. Thomas, P. Cheiney,
  and L. Tarruell,
  Quantum liquid droplets in a mixture of Bose-Einstein condensates,
  Science \textbf{359}, 301 (2018).

\bibitem{Semeghini}
  G. Semeghini, G. Ferioli, L. Masi, C. Mazzinghi, L. Wolswijk,
  F. Minardi, M. Modugno, G. Modugno, M. Inguscio, and M. Fattori,
  Self-bound quantum droplets of atomic mixtures in free space,
  Phys. Rev. Lett. \textbf{120}, 235301 (2018).

\bibitem{Cheiney}
  P. Cheiney, C. R. Cabrera, J. Sanz, B. Naylor, L. Tanzi, and L. Tarruell,
  Bright soliton to quantum droplet transition in a mixture of
  Bose-Einstein condensates,
  Phys. Rev. Lett. \textbf{120}, 135301 (2018).

\bibitem{waechtler2016}
  F. W\"achtler and L. Santos,
  Quantum filaments in dipolar Bose-Einstein condensates,
  Phys. Rev. A \textbf{93}, 061603(R) (2016).

\bibitem{waechtler2016b}
  F. W\"achtler and L. Santos,
  Ground-state properties and elementary excitations of quantum
  droplets in dipolar Bose-Einstein condensates,
  Phys. Rev. A \textbf{94}, 043618 (2016).

\bibitem{Saito}
  H. Saito,
  Path-integral Monte Carlo study on a droplet of a dipolar
  Bose-Einstein condensate stabilized by quantum fluctuation,
  J. Phys. Soc. Jpn. \textbf{85}, 053001 (2016).

\bibitem{Bisset}
  R. N. Bisset, R. M. Wilson, D. Baillie, and P. B. Blakie,
  Ground-state phase diagram of a dipolar condensate with quantum
  fluctuations,
  Phys. Rev. A \textbf{94}, 033619 (2016).

\bibitem{macia2016droplets}
  A. Macia, J. S\'anchez-Baena, J. Boronat, and F. Mazzanti,
  Droplets of trapped quantum dipolar bosons,
  Phys. Rev. Lett. \textbf{117}, 205301 (2016).

\bibitem{8PhysRevA.96.053630}
  M. Wenzel, F. B\"ottcher, T. Langen, I. Ferrier-Barbut, and T. Pfau,
  Striped states in a many-body system of tilted dipoles,
  Phys. Rev. A \textbf{96}, 053630 (2017).

\bibitem{9PhysRevLett.119.255302}
  D. Baillie, R. M. Wilson, and P. B. Blakie,
  Collective excitations of self-bound droplets of a dipolar quantum fluid,
  Phys. Rev. Lett. \textbf{119}, 255302 (2017).

\bibitem{cidrim2018vortices}
  A. Cidrim, F. E. A. dos Santos, E. A. L. Henn, and T. Macr\'i,
  Vortices in self-bound dipolar droplets,
  Phys. Rev. A \textbf{98}, 023618 (2018).

\bibitem{13PhysRevA.98.063602}
  Y. Li, Z. Chen, Z. Luo, C. Huang, H. Tan, W. Pang, and B. A. Malomed,
  Two-dimensional vortex quantum droplets,
  Phys. Rev. A \textbf{98}, 063602 (2018).

\bibitem{14PhysRevLett.122.193902}
  Y. V. Kartashov, B. A. Malomed, and L. Torner,
  Metastability of quantum droplet clusters,
  Phys. Rev. Lett. \textbf{122}, 193902 (2019).

\bibitem{15PhysRevResearch.1.033088}
  F. B\"ottcher, M. Wenzel, J.-N. Schmidt, M. Guo, T. Langen,
  I. Ferrier-Barbut, T. Pfau, R. Bomb\'in, J. S\'anchez-Baena,
  J. Boronat, and F. Mazzanti,
  Dilute dipolar quantum droplets beyond the extended Gross-Pitaevskii
  equation,
  Phys. Rev. Res. \textbf{1}, 033088 (2019).

\bibitem{16PhysRevLett.122.090401}
  G. Ferioli, G. Semeghini, L. Masi, G. Giusti, G. Modugno, M. Inguscio,
  A. Gallem\'i, A. Recati, and M. Fattori,
  Collisions of self-bound quantum droplets,
  Phys. Rev. Lett. \textbf{122}, 090401 (2019).

\bibitem{17PhysRevLett.123.133901}
  X. Zhang, X. Xu, Y. Zheng, Z. Chen, B. Liu, C. Huang, B. A. Malomed,
  and Y. Li,
  Semidiscrete quantum droplets and vortices,
  Phys. Rev. Lett. \textbf{123}, 133901 (2019).

\bibitem{oldziejewski2020strongly}
  R. O\l dziejewski, W. G\'orecki, K. Paw\l owski, and K. Rz\k{a}\.{z}ewski,
  Strongly correlated quantum droplets in quasi-1D dipolar Bose gas,
  Phys. Rev. Lett. \textbf{124}, 090401 (2020).

\bibitem{30bisset2021}
  R. N. Bisset, L. A. Pe\~{n}a Ardila, and L. Santos,
  Quantum droplets of dipolar mixtures,
  Phys. Rev. Lett. \textbf{126}, 025301 (2021).

\bibitem{smith2021quantum}
  J. C. Smith, D. Baillie, and P. B. Blakie,
  Quantum droplet states of a binary magnetic gas,
  Phys. Rev. Lett. \textbf{126}, 025302 (2021).

\bibitem{Lee}
  A.-C. Lee, D. Baillie, and P. B. Blakie,
  Numerical calculation of dipolar-quantum-droplet stationary states,
  Phys. Rev. Res. \textbf{3}, 013283 (2021).

\bibitem{Yogurt}
  T. A. Yo\u{g}urt, A. Kele\c{s}, and M. \"{O}. Oktel,
  Spinor boson droplets stabilized by spin fluctuations,
  Phys. Rev. A \textbf{105}, 043309 (2022).

\bibitem{20PhysRevLett.130.043401}
  J. Kopyci\'{n}ski, M. \L ebek, W. G\'orecki, and K. Paw\l owski,
  Ultrawide dark solitons and droplet-soliton coexistence in
  a dipolar Bose gas with strong contact interactions,
  Phys. Rev. Lett. \textbf{130}, 043401 (2023).

%spinor dipolar BEC
\bibitem{22PhysRevLett.93.040403}
  S. Yi, L. You, and H. Pu,
  Quantum phases of dipolar spinor condensates,
  Phys. Rev. Lett. \textbf{93}, 040403 (2004).

\bibitem{Kawaguchi_EdH}
  Y. Kawaguchi, H. Saito, and M. Ueda,
  Einstein-de Haas effect in dipolar Bose-Einstein condensates,
  Phys. Rev. Lett. \textbf{96}, 080405 (2006).

\bibitem{Santos}
  L. Santos and T. Pfau,
  Spin-3 chromium Bose-Einstein condensates,
  Phys. Rev. Lett. \textbf{96}, 190404 (2006).

\bibitem{Diener}
  R. B. Diener and T.-L. Ho,
  $^{52}{\rm Cr}$ spinor condensate: a biaxial or uniaxial spin nematic,
  Phys. Rev. Lett. \textbf{96}, 190405 (2006).

\bibitem{Yi}
  S. Yi and H. Pu,
  Spontaneous spin textures in dipolar spinor condensates,
  Phys. Rev. Lett. \textbf{97}, 020401 (2006).

\bibitem{Kawaguchi2}
  Y. Kawaguchi, H. Saito, and M. Ueda,
  Spontaneous circulation in ground-state spinor dipolar Bose-Einstein
  condensates,
  Phys. Rev. Lett. \textbf{97}, 130404 (2006).

\bibitem{26PhysRevLett.98.110406}
  Y. Kawaguchi, H. Saito, and M. Ueda,
  Can spinor dipolar effects be observed in Bose-Einstein condensates?,
  Phys. Rev. Lett. \textbf{98}, 110406 (2007).

\bibitem{Takahashi}
  M. Takahashi, S. Ghosh, T. Mizushima, and K. Machida,
  Spinor dipolar Bose-Einstein condensates: Classical spin approach,
  Phys. Rev. Lett. \textbf{98}, 260403 (2007).

\bibitem{Gawryluk}
  K. Gawryluk, M. Brewczyk, K. Bongs, and M. Gajda,
  Resonant Einstein-de Haas effect in a rubidium condensate,
  Phys. Rev. Lett. \textbf{99}, 130401 (2007).

\bibitem{Sun}
  B. Sun and L. You,
  Observing the Einstein-de Haas effect with atoms in an optical lattice,
  Phys. Rev. Lett. \textbf{99}, 150402 (2007).

\bibitem{Yasunaga}
  M. Yasunaga and M. Tsubota,
  Spin echo in spinor dipolar Bose-Einstein condensates,
  Phys. Rev. Lett. \textbf{101}, 220401 (2008).

\bibitem{hoshi2010}
  S. Hoshi and H. Saito,
  Symmetry-breaking magnetization dynamics of spinor dipolar Bose-Einstein condensates,
  Phys. Rev. A \textbf{81}, 013627 (2010).

\bibitem{Swislocki}
  T. \'Swis\l ocki, M. Brewczyk, M. Gajda, and K. Rz\k{a}\.{z}ewski,
  Spinor condensate of $^{87}{\rm Rb}$ as a dipolar gas,
  Phys. Rev. A \textbf{81}, 033604 (2010).

\bibitem{Huhtamaki}
  J. A. M. Huhtam\"aki, M. Takahashi, T. P. Simula, T. Mizushima, and
  K. Machida,
  Spin textures in condensates with large dipole moments,
  Phys. Rev. A \textbf{81}, 063623 (2010).

\bibitem{Deure}
  F. Deuretzbacher, G. Gebreyesus, O. Topic, M. Scherer, B. L\"ucke,
  W. Ertmer, J. Arlt, C. Klempt, and L. Santos,
  Parametric amplification of matter waves in dipolar spinor
  Bose-Einstein condensates,
  Phys. Rev. A \textbf{82}, 053608 (2010).

\bibitem{kudo2010hydrodynamic}
  K. Kudo and Y. Kawaguchi,
  Hydrodynamic equation of a spinor dipolar Bose-Einstein condensate,
  Phys. Rev. A \textbf{82}, 053614 (2010).

\bibitem{Simula}
  T. P. Simula, J. A. M. Huhtam\"aki, M. Takahashi, T. Mizushima, and
  K. Machida,
  Rotating dipolar spin-1 Bose-Einstein condensates,
  J. Phys. Soc. Jpn. \textbf{80}, 013001 (2011).

\bibitem{Gaw11}
  K. Gawryluk, K. Bongs, and M. Brewczyk,
  How to observe dipolar effects in spinor Bose-Einstein condensates,
  Phys. Rev. Lett. \textbf{106}, 140403 (2011).

\bibitem{Pasquiou}
  B. Pasquiou, E. Mar\'echal, G. Bismut, P. Pedri, L. Vernac,
  O. Gorceix, and B. Laburthe-Tolra,
  Spontaneous demagnetization of a dipolar spinor Bose gas in an
  ultralow magnetic field,
  Phys. Rev. Lett. \textbf{106}, 255303 (2011).

\bibitem{Eto}
  Y. Eto, H. Saito, and T. Hirano,
  Observation of dipole-induced spin texture in an $^{87}{\rm Rb}$
  Bose-Einstein condensate,
  Phys. Rev. Lett. \textbf{112}, 185301 (2014).

\bibitem{Marti}
  G. E. Marti, A. MacRae, R. Olf, S. Lourette, F. Fang, and
  D. M. Stamper-Kurn,
  Coherent magnon optics in a ferromagnetic spinor Bose-Einstein
  condensate,
  Phys. Rev. Lett. \textbf{113}, 155302 (2014).

\bibitem{Kunimi}
  H. Saito and M. Kunimi,
  Energy shift of magnons in a ferromagnetic spinor-dipolar
  Bose-Einstein condensate,
  Phys. Rev. A \textbf{91}, 041603(R) (2015).

\bibitem{Oshima}
  T. Oshima and Y. Kawaguchi,
  Spin Hall effect in a spinor dipolar Bose-Einstein condensate,
  Phys. Rev. A \textbf{93}, 053605 (2016).

\bibitem{Borgh}
  M. O. Borgh, J. Lovegrove, and J. Ruostekoski,
  Internal structure and stability of vortices in a dipolar spinor
  Bose-Einstein condensate,
  Phys. Rev. A \textbf{95}, 053601 (2017).

\bibitem{liao2017anisotropic}
  B. Liao, S. Li, C. Huang, Z. Luo, W. Pang, H. Tan, B. A. Malomed,
  and Y. Li,
  Anisotropic semivortices in dipolar spinor condensates controlled by
  Zeeman splitting,
  Phys. Rev. A \textbf{96}, 043613 (2017).

\bibitem{lepoutre2018spin}
  S. Lepoutre, K. Kechadi, B. Naylor, B. Zhu, L. Gabardos, L. Isaev, P. Pedri, A. M. Rey, L. Vernac, and B. Laburthe-Tolra,
  Spin mixing and protection of ferromagnetism in a spinor dipolar condensate,
  Phys. Rev. A \textbf{97}, 023610 (2018).

\bibitem{33PhysRevLett.129.223401}
  Y. Miyazawa, R. Inoue, H. Matsui, G. Nomura, and M. Kozuma,
  Bose-Einstein condensation of europium,
  Phys. Rev. Lett. \textbf{129}, 223401 (2022).

\bibitem{34PhysRevA.81.041603}
  D. J. Papoular, G. V. Shlyapnikov, and J. Dalibard,
  Microwave-induced Fano-Feshbach resonances,
  Phys. Rev. A \textbf{81}, 041603(R) (2010).

\bibitem{35JCP}
  A. A. Buchachenko, G. Cha\l asi\'nski, and M. M. Szcz\c{e}\'sniak,
  Europium dimer: van der Waals molecule with extremely weak antiferromagnetic spin coupling,
  J. Chem. Phys. \textbf{131}, 241102 (2009).

\bibitem{36PhysRevA.81.022701}
  Y. V. Suleimanov,
  Zeeman relaxation of magnetically trapped Eu atoms,
  Phys. Rev. A \textbf{81}, 022701 (2010).

\bibitem{lima2011}
  A. R. P. Lima and A. Pelster,
  Quantum fluctuations in dipolar Bose gases,
  Phys. Rev. A \textbf{84}, 041604(R) (2011).

\bibitem{lima2012}
  A. R. P. Lima and A. Pelster,
  Beyond mean-field low-lying excitations of dipolar Bose gases,
  Phys. Rev. A \textbf{86}, 063609 (2012).

\bibitem{43jin2002}
  W.-G. Jin, T. Endo, T. Wakui, H. Uematsu, T. Minowa, and H. Katsuragawa,
  Measurements of the Hyperfine Structure and $\Delta F=+2$ Transitions in Eu I by High-Resolution Diode-Laser Spectroscopy,
  J. Phys. Soc. Jpn. \textbf{71}, 1905 (2002).

\bibitem{Richardson}
  O. W. Richardson,
  A mechanical effect accompanying magnetization,
  Phys. Rev. Ser. I \textbf{26}, 248 (1908).

\bibitem{EdH}
  A. Einstein and W. J. de Haas,
  Experimental proof of Amp\'ere's molecular currents,
  Verh. Dtsch. Phys. Ges. \textbf{17}, 152 (1915).

\bibitem{StamperKurn}
  L. E. Sadler, J. M. Higbie, S. R. Leslie, M. Vengalattore, and
  D. M. Stamper-Kurn,
  Spontaneous symmetry breaking in a quenched ferromagnetic spinor
  Bose-Einstein condensate,
  Nature (London) \textbf{443}, 312 (2006).

%supersolid Experiment
%\bibitem{Donner}
%J. L\'eonard, A. Morales, P. Zupancic, T. Esslinger, and T. Donner,
%Supersolid formation in a quantum gas breaking a continuous
%translational symmetry,
%Nature (London) \textbf{543}, 87 (2017).

%\bibitem{Li}
%  J.-R. Li, J. Lee, W. Huang, S. Burchesky, B. Shteynas,
%  F. \c{C}. Top, A. O. Jamison, and W. Ketterle,
%  A stripe phase with supersolid properties in spin-orbit-coupled
%  Bose-Einstein condensates,
%  Nature (London) \textbf{543}, 91 (2017).

\bibitem{Tanzi19}
L. Tanzi, S. M. Roccuzzo, E. Lucioni, F. Fam\'a, A. Fioretti,
C. Gabbanini, G. Modugno, A. Recati, and S. Stringari,
Supersolid symmetry breaking from compressional oscillations in a
dipolar quantum gas,
Nature (London) \textbf{574}, 382 (2019).

\bibitem{Bottcher19}
F. B\"ottcher, J.-N. Schmidt, M. Wenzel, J. Hertkorn, M. Guo,
T. Langen, and T. Pfau,
Transient supersolid properties in an array of dipolar quantum droplets,
Phys. Rev. X \textbf{9}, 011051 (2019).

\bibitem{Chomaz19}
L. Chomaz, D. Petter, P. Ilzh\"ofer, G. Natale, A. Trautmann,
C. Politi, G. Durastante, R. M. W. van Bijnen, A. Patscheider,
M. Sohmen, M. J. Mark, and F. Ferlaino, 
Long-lived and transient supersolid behaviors in dipolar quantum gases,
Phys. Rev. X \textbf{9}, 021012 (2019).

\bibitem{Tanzi19_2}
L. Tanzi, S. M. Roccuzzo, E. Lucioni, F. Fam\'a, A. Fioretti,
C. Gabbanini, G. Modugno, A. Recati, and S. Stringari,
Supersolid symmetry breaking from compressional oscillations in a
dipolar quantum gas,
Nature (London) \textbf{574}, 382 (2019).

\bibitem{Guo19}
M. Guo, F. B\"ottcher, J. Hertkorn, J.-N. Schmidt, M. Wenzel,
H. P. B\"uchler, T. Langen, and T. Pfau,
The low-energy Goldstone mode in a trapped dipolar supersolid,
Nature (London) \textbf{574}, 386 (2019).

\bibitem{Natale19}
G. Natale, R. M. W. van Bijnen, A. Patscheider, D. Petter, M. J. Mark,
L. Chomaz, and F. Ferlaino,
Excitation spectrum of a trapped dipolar supersolid and its
experimental evidence,
Phys. Rev. Lett. \textbf{123}, 050402 (2019).

\bibitem{Tanzi21}
L. Tanzi, J. G. Maloberti, G. Biagioni, A. Fioretti, C. Gabbanini, and
G. Modugno,
Evidence of superfluidity in a dipolar supersolid from nonclassical
rotational inertia,
Science \textbf{371}, 1162 (2021).

\bibitem{Ilzhofer21}
P. Ilzh\"ofer, M. Sohmen, G. Durastante, C. Politi, A. Trautmann,
G. Natale, G. Morpurgo, T. Giamarchi, L. Chomaz, M. J. Mark, and
F. Ferlaino,
Phase coherence in out-of-equilibrium supersolid states of ultracold
dipolar atoms,
Nat. Phys. \textbf{17}, 356 (2021).

\bibitem{Hertkorn21}
J. Hertkorn, J.-N. Schmidt, F. B\"ottcher, M. Guo, M. Schmidt,
K. S. H. Ng, S. D. Graham, H. P. B\"uchler, T. Langen, M. Zwierlein,
and T. Pfau, 
Density fluctuations across the superfluid-supersolid phase transition
in a dipolar quantum gas,
Phys. Rev. X \textbf{11}, 011037 (2021).

\bibitem{Sohmen21}
M. Sohmen, C. Politi, L. Klaus, L. Chomaz, M. J. Mark, M. A. Norcia,
and F. Ferlaino,
Birth, life, and death of a dipolar supersolid,
Phys. Rev. Lett. \textbf{126}, 233401 (2021).

\bibitem{Norcia21}
M. A. Norcia, C. Politi, L. Klaus, E. Poli, M. Sohmen, M. J. Mark,
R. N. Bisset, L. Santos, and F. Ferlaino,
Two-dimensional supersolidity in a dipolar quantum gas,
Nature (London) \textbf{596}, 357 (2021).

\bibitem{Bland22}
T. Bland, E. Poli, C. Politi, L. Klaus, M. A. Norcia, F. Ferlaino,
L. Santos, and R. N. Bisset,
Two-dimensional supersolid formation in dipolar condensates,
Phys. Rev. Lett. \textbf{128}, 195302 (2022).

\bibitem{Biagioni22}
G. Biagioni, N. Antolini, A. Ala\~{n}a, M. Modugno, A. Fioretti,
C. Gabbanini, L. Tanzi, and G. Modugno,
Dimensional crossover in the superfluid-supersolid quantum phase transition,
Phys. Rev. X \textbf{12}, 021019 (2022).

\bibitem{Norcia22}
M. A. Norcia, E. Poli, C. Politi, L. Klaus, T. Bland, M. J. Mark,
L. Santos, R. N. Bisset, and F. Ferlaino,
Can angular oscillations probe superfluidity in dipolar supersolids?,
Phys. Rev. Lett. \textbf{129}, 040403 (2022).

\bibitem{Sanchez23}
  J. S\'anchez-Baena, C. Politi, F. Maucher, F. Ferlaino, and T. Pohl,
  Heating a dipolar quantum fluid into a solid,
  Nat. Comm. \textbf{14}, 1868 (2023).

\end{thebibliography}
\end{document}